\documentclass{moriond}

\usepackage{tikz}
\usepackage{tikz-feynman}
\usepackage{subcaption}
\usepackage{adjustbox}
\usepackage{caption}
\def\be{\begin{equation}}
\def\ee{\end{equation}}
\def\bea{\begin{eqnarray}}
\def\eea{\end{eqnarray}}

\newcommand{\Photo}{}

\begin{document}
\vspace*{4cm}
\title{$b\to s\ell^+\ell^- (\ell = e, \mu, \tau)$ and $b\to s\nu\bar\nu$ at Belle and Belle II}

\author{Meihong Liu\\ on behalf of the Belle and Belle II collaborations}

\address{Jilin University, Changchun 130012, People’s Republic of China;\\
Deutsches Elektronen–Synchrotron, Hamburg 22607, Germany.}

\maketitle\abstracts{
%This is where the abstract should be placed. It should consist of one paragraph
%and give a concise summary of the material in the article below.
%Replace the title, authors, and addresses within the curly brackets
%with your own title, authors, and addresses; please use
%capital letters for the title and the authors. You may have as many authors and
%addresses as you wish. It's preferable not to use footnotes in the abstract
%or the title; the acknowledgments for funding bodies etc. are placed in a separate section at
%the end of the text\\
The Belle and Belle~II experiments have accumulated a combined data set of $1.2~\mathrm{ab}^{-1}$ of $e^+e^- \to B\bar{B}$ collisions at the $\Upsilon(4S)$ resonance. Owing to the clean event environment and the well-constrained initial-state kinematics, these data are particularly well suited for studying channels involving missing energy from neutrinos. This includes the reinterpretation of $B^+\to K^+\nu\bar{\nu}$ and the first inclusive measurement of $B\to X_s\nu\bar{\nu}$ with the missing energy directly from $B$ decays, as well as searches involving missing energy from $\tau$ decays, such as $B^0\to K^{*0}\tau^+\tau^-$, $B^0\to K_S^{0}\tau^+\tau^-$, and $B^+\to K^{+}\tau^+\tau^-$.}

\section{Introduction}

\begin{figure}[htp]
    \centering
    \begin{subfigure}[b]{0.25\linewidth}
        \centering
        \includegraphics[width=\linewidth]{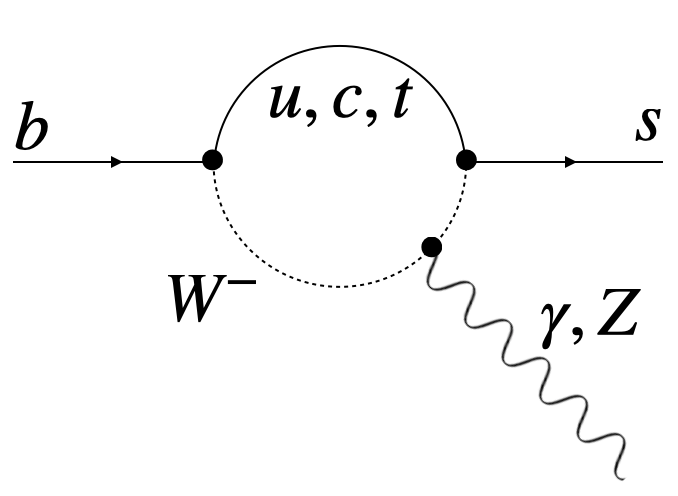}
        \caption{SM contributions}
        \label{fig:diagram1}
    \end{subfigure}
    \hfill
    \begin{subfigure}[b]{0.25\linewidth}
        \centering
        \includegraphics[width=\linewidth]{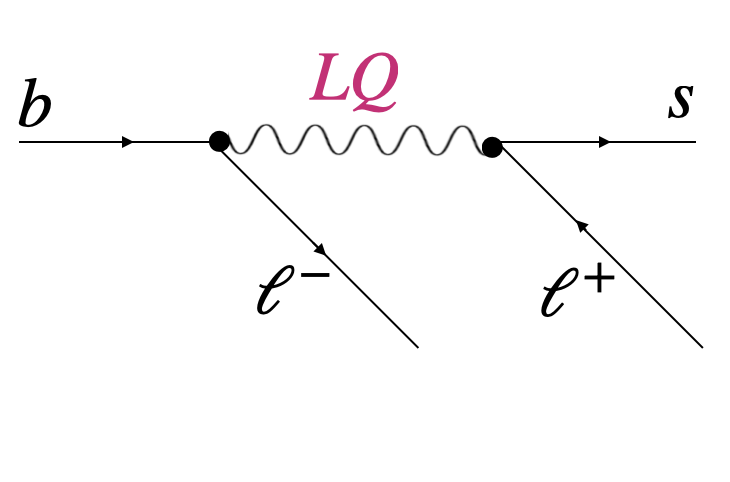}
        \caption{Leptoquark}
        \label{fig:diagram2}
    \end{subfigure}
    \hfill
    \begin{subfigure}[b]{0.25\linewidth}
        \centering
        \includegraphics[width=\linewidth]{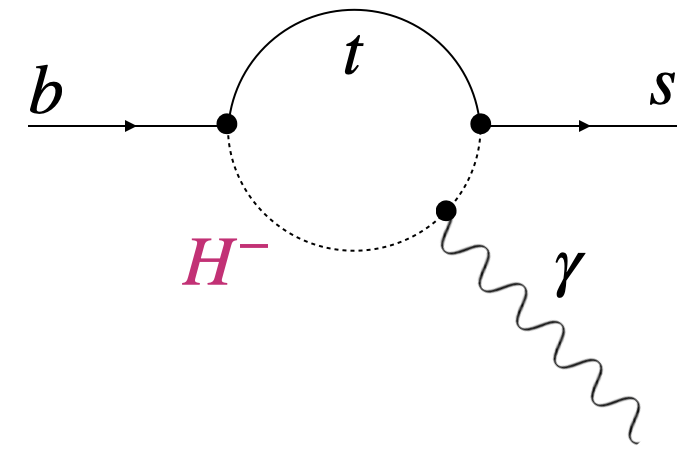}
        \caption{Charged Higgs}
        \label{fig:diagram3}
    \end{subfigure}
    \caption{Diagrams representing $b\to s$ transitions: Example of SM contributions (a) and possible NP scenarios (b, c).}
    \label{fig:diagram}
\end{figure}

The quark transition $b\to s$, mediated by a flavour-changing neutral current (FCNC), is a sensitive probe of physics beyond the Standard Model (SM). In the SM, FCNC processes occur only through loop diagrams such as the electroweak penguin amplitudes in Fig.~\ref{fig:diagram1}, leading to small branching fractions of $\mathcal{O}(10^{-7})$--$\mathcal{O}(10^{-5})$. Searches at Belle and Belle~II aim to reveal possible New Physics (NP) effects via modified rates, either from new tree-level interactions (Fig.~\ref{fig:diagram2}) or reduced GIM suppression in loops (Fig.~\ref{fig:diagram3}).

The Belle experiment~\cite{bdetector} operated at the KEKB asymmetric $e^+e^-$ collider~\cite{kekb} in Tsukuba, Japan. Its successor, Belle~II~\cite{b2detector}, runs at SuperKEKB, designed to reach about forty times higher instantaneous luminosity~\cite{superkekb}. (Super)KEKB collides $e^+e^-$ at the $\Upsilon(4S)$ resonance, producing almost exclusively $B\bar B$ pairs in a low-background environment compared to hadron colliders. The data samples used here correspond to 711\,fb$^{-1}$ from Belle and 365\,fb$^{-1}$ from Belle~II. Since $\Upsilon(4S)\to B\bar B$ produces no additional particles, missing energy from neutrinos in the signal decay can be inferred from the reconstruction of the accompanying $B$ meson ($B_{\rm tag}$). The large solid-angle coverage and well-constrained initial kinematics of Belle and Belle~II make it particularly well suited for analyses with missing energy, using the two tagging methods described below.

~

\noindent\textbf{Inclusive tagging}: 
%\begin{itemize}
%\item
this method exploits the inclusive properties of the accompanying $B$ decay by exploiting distinct signal features with machine-learning. It offers high efficiency, but suffers from large background and a strong reliance on simulation.
%\end{itemize}

~

\noindent\textbf{Exclusive tagging:}
This approach employs the machine-learning-based Full Event Interpretation (FEI) algorithm~\cite{fei} to reconstruct the tag-side $B$ meson using either hadronic or semileptonic decay modes. By applying a stringent requirement on the FEI output probability, a high-purity sample can be obtained.

\begin{itemize}
\item \textbf{Hadronic tagging} fully reconstructs specific hadronic $B$ decays. It strongly suppresses combinatorial background and provides tight kinematic constraints on $B_{\rm sig}$ via $\vec{p}_{B_{\rm sig}} = -\vec{p}_{B_{\rm tag}}$ in the $\Upsilon(4S)$ rest frame. However, the reconstruction efficiency is typically below 1\%.

\item \textbf{Semileptonic tagging} reconstructs $B\to D\ell\nu$ and $B\to D^*\ell\nu$ decays, where the neutrino is treated as missing energy. The presence of a high-momentum lepton enables efficient identification, resulting in higher efficiency than hadronic tagging due to large semileptonic branching fractions. However, the undetected neutrino leads to reduced kinematic constraints and lower sample purity.
\end{itemize}

\smallskip
This report is organized as follows. Sections~\ref{sec:kstar0tauatu}, \ref{sec:kstauatu}, and \ref{sec:ktauatu} present the latest results on $b \to s \ell \ell$ transitions in $B^0\to K^{*0}\tau^+\tau^-$ ~\cite{ktt}, $B^0\to K_S^{0}\tau^+\tau^-$, and $B^+\to K^{+}\tau^+\tau^-$ \cite{kptt} decays, respectively. Sections~\ref{KNN} and \ref{SNN} discuss the reinterpretation of $B^+ \to K^+ \nu \bar{\nu}$ evidence~\cite{knn_reint} and the search for inclusive $B \to X_s \nu \bar{\nu}$ decay~\cite{svv}.

\section{$b\to s\ell^+\ell^-$ ($b\to s\tau^+\tau^-$) transitions}\label{sec:sellell}

%Belle~II offers complementary information to the extensively studied $b \to s e^+e^-$ and $b \to s \mu^+\mu^-$ transitions. Accessing distinctive features in these processes requires a substantially larger data set than currently available from the combined Belle and Belle~II samples. % Nonetheless, a recent measurement of the inclusive $B \to X_s \ell^+\ell^-$ decay has shown intriguing results.
In this report, we concentrate on final states containing $\tau$-lepton pairs. 
The $b \to s \tau^+ \tau^-$ processes involve third-generation leptons, which makes them particularly sensitive to  potential new physics contributions~\cite{ktt_theory}. Furthermore, the ability of Belle~II to reconstruct final states with undetectable neutrinos from $\tau$ decays provides a significant experimental advantage.

\subsection{$B^0\to K^{*0} \tau^+ \tau^-$ search using hadronic $B$-tagging at Belle II }\label{sec:kstar0tauatu}

Signal candidates are categorized into $\ell\ell$, $\ell\pi$, $\pi\pi$, and $\rho X$ categories according to the $\tau$ decay products ($t_\tau$). The excellent photon efficiency and resolution enable a clean reconstruction of $\pi^0$ mesons, allowing $\rho^+\to \pi^+\pi^0$ decays to be included as $\tau$ prong candidates.

\begin{figure}
\centerline{\includegraphics[width=0.9\linewidth]{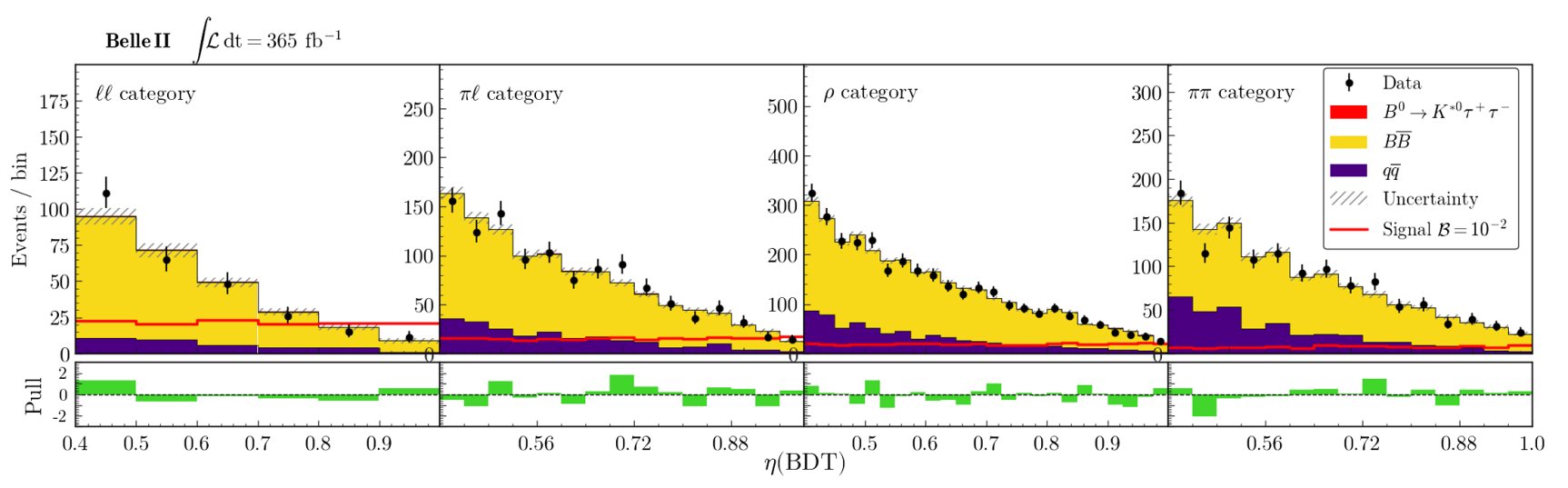}}
\caption[]{Distributions of $\eta$(BDT) in signal region for the four signal categories. 
    The signal, fitted with $\mathcal{B} = [-0.15 \pm 1.01] \times 10^{-3}$ and scaled assuming $\mathcal{B} =10^{-2}$, is shown as a reference.}
\label{fig:kstartautau}
\end{figure}

Among these, the $\ell\ell$ category provides the highest sensitivity and purity due to the excellent lepton identification at Belle~II and the strong suppression of $q\bar{q}$ background. The latter originates from the fact that continuum $q\bar{q}$ events hadronize into jet-like topologies dominated by light hadrons, resulting in event shapes that differ significantly from the more spherical $B\bar B$ topology. A boosted decision tree (BDT) is trained to further suppress background using variables such as residual calorimeter energy in the electromagnetic calorimeter ($E_{\rm extra}$), the $K^{*0}t_\tau$ invariant mass, $q^2$, etc. A simultaneous fit to the BDT outputs (Fig.~\ref{fig:kstartautau}) shows no significant signal. We obtain a limit of $\mathcal{B}^{\mathrm{UL}} >1.8\times10^{-3}$ at 90\% confidence limit (CL), which improves the Belle constraint by approximately a factor of two while using only half of its data.

%\begin{figure}[htbp]
   % \centering 
    %\includegraphics[width=0.9\linewidth]{ktt.png}
    %\caption{Distributions of $\eta$(BDT) in signal region for the four signal categories. The signal, fitted with $\mathcal{B} = [-0.15 \pm 1.01] \times 10^{-3}$ and scaled assuming $\mathcal{B} =10^{-2}$, is shown as a reference. %The bottom panel shows the pull distributions.
     %}
    %\label{fig:kstartautau}
%\end{figure}

\subsection{First $B^0\to K_S^0 \tau^+ \tau^-$ search using hadronic $B$-tagging at Belle and Belle II}\label{sec:kstauatu}
\begin{figure}[htbp]
    \centering 
    \includegraphics[width=0.7\linewidth]{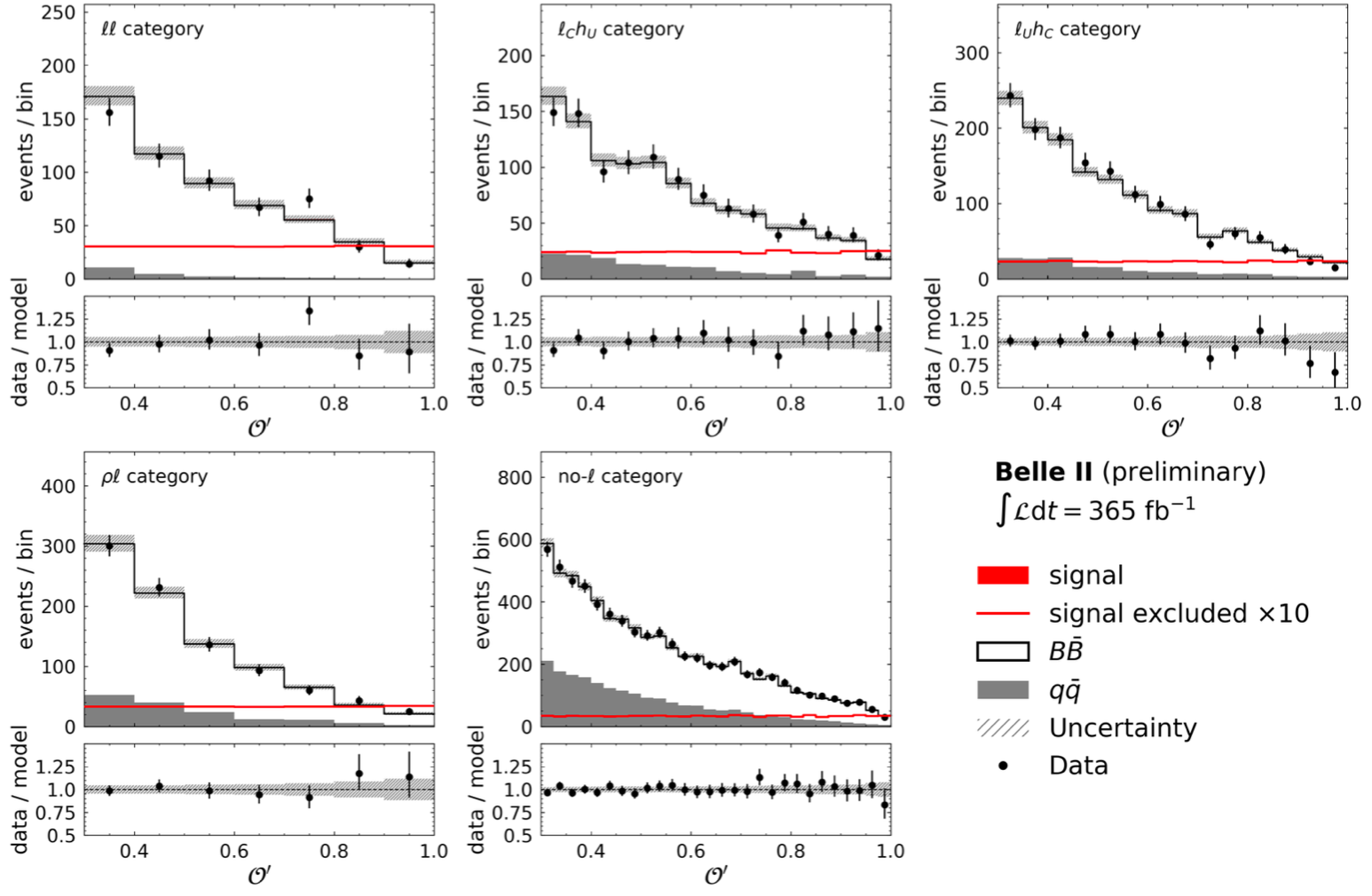}
    \caption{Post-fit distributions of $\mathcal{O}^\prime$ for Belle II (as example). The red filled histogram represents the measured $\mathcal{B}(B^0\to K_S^0\tau^+\tau^-)$, red step histogram shows the signal $\mathcal{B}$ excluded at the obtained upper limit multiplied by 10.%, together with the simulated $B\bar B$ background contributions (open histogram) and $q\bar q$ background contributions (gray filled histogram).  
    %The $B^0 \rightarrow K^{*0} \tau^+ \tau^-$ signal, fitted with a branching fraction of $[-0.15 \pm 1.01] \times 10^{-3}$ and scaled assuming a branching fraction of $10^{-2}$, is shown as a reference. The bottom panel shows the pull distributions. 
    }
    \label{fig:kstautau}
\end{figure}

Signal candidates are divided into $\ell\ell$, $\ell_c h_u$, $\ell_u h_c$, $\rho\ell$, and no-$\ell$ categories to account for different background compositions and purities. BDTs are used for background suppression based on variables such as $E_{\rm extra}$, invariant-mass combinations of $K_S^0$ and $\tau$-prong, $\tau$-prong kinematics, event-shape observables, and $q^2$, etc. A simultaneous fit to the transformed BDT output $\mathcal{O}'$ across all categories and datasets (Fig.~\ref{fig:kstautau}) show no significant signal. An upper limit of $\mathcal{B}^{\mathrm{UL}} < 8.4\times10^{-4}$ at 90\% CL is set, representing the first search result on the $B^0\to K_S^0 \tau^+ \tau^-$ decay.

\subsection{$B^+\to K^+ \tau^+ \tau^-$ search using hadronic $B$-tagging at Belle and Belle II}\label{sec:ktauatu}

%Belle and Belle II perform a combined search for $B^+ \to K^+ \tau^+\tau^-$ using hadronic $B$-tagging. 
On the signal side, only leptonic $\tau$ decays ($\tau \to e\nu\bar{\nu}$ and $\tau \to \mu\nu\bar{\nu}$) are considered, providing a low-background environment. Instead of a multivariate approach, a cut-based strategy is adopted: the dominant $B \to D\ell\nu$ background is suppressed by requiring the $K\ell$ invariant mass to exceed the $D$-meson threshold, the lepton momentum, and the $q^2$.

The signal yield is extracted from the distribution of $E_{\rm extra}$, using a Poisson event-counting method optimized in a signal-enhanced region. Signal peaks at 0 in $E_{\rm extra}$ as it doesn't have extra photons in the calorimeter and backrgound distributes far from 0. The background expectation is validated with sideband data. No significant excess is observed, and an upper limit of $\mathcal{B}(B^+ \to K^+ \tau^+\tau^-) < 5.6 \times 10^{-4}$ at 90\% CL is set (the upper limit has been updated compared to the result presented at the conference \cite{kptt}), representing the most stringent constraint on this channel to date.

\begin{figure}[htbp]
    \centering 
    \includegraphics[width=0.8\linewidth]{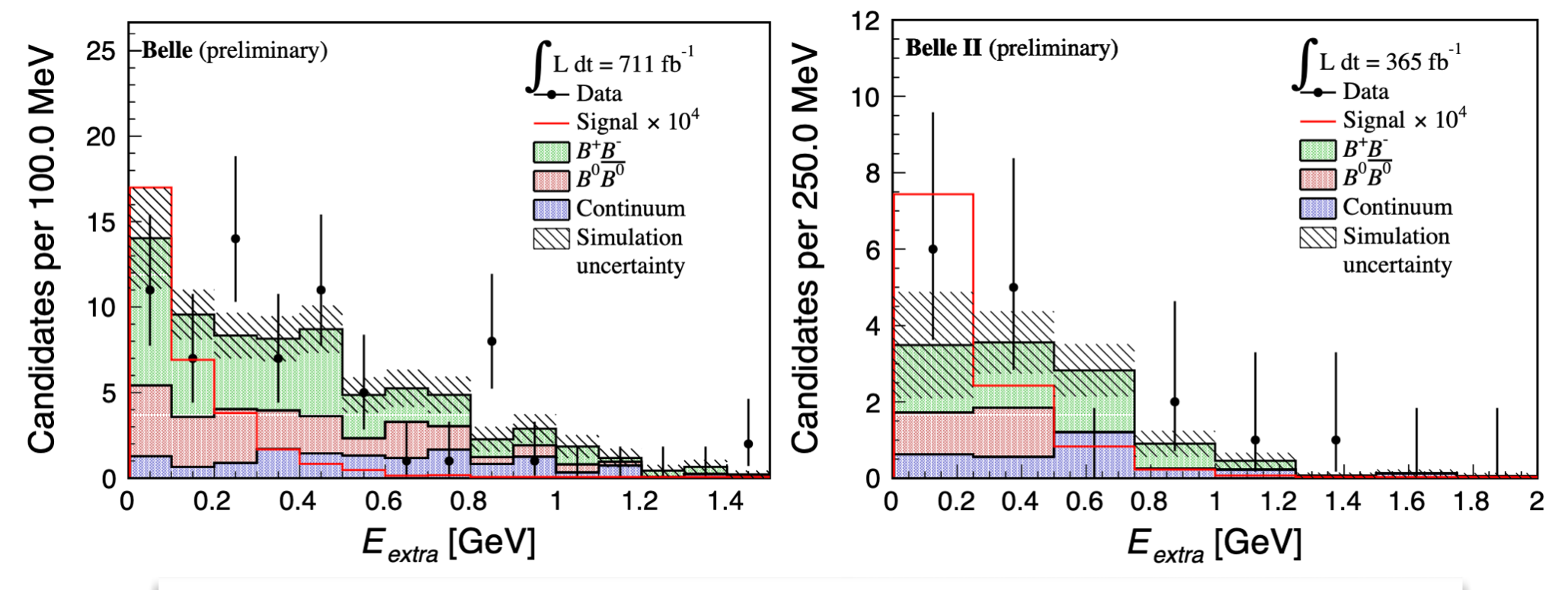}
    \caption{Distributions of $E_{\mathrm{extra}}$ after final selections in (left) Belle and (right) Belle~II data. Expectations based on corrected simulation for backgrounds and for $10^4\times$-enhanced SM signal are overlaid.}
    \label{fig:ktautau}
\end{figure}

\section{$b\to s\nu\bar\nu$ transitions}\label{sec:svv}

\subsection{$B^+\to K^+\nu\bar\nu$ reinterpretation}\label{KNN}

We recently determined of the $\mathcal{B}(B^{+}\to K^{+}\nu\bar{\nu})$ decay at Belle~II under the assumption of SM kinematics, providing the first evidence fo this decay using combined inclusive and hadronic $B$-tagging method~\cite{knn}. 
In this analysis, backgrounds are suppressed using two consecutive classifiers: BDT$_1$ and BDT$_2$. The signal yield is extracted using a two-dimensional binning scheme in the transformed BDT$_2$ output, $\eta_{\mathrm{BDT_2}}$, and the reconstructed momentum-transfer squared, $q^2_{\mathrm{rec}}$, which is divided into $4\times 3$ bins.

The reinterpretation strategy is straightforward~\cite{reint}. We start with simulated SM signal events and construct a model-agnostic number density, $n_0(x)$, where $x$ denotes the fit observables, namely $\eta_{\mathrm{BDT_2}}$ and $q^2_{\rm rec}$ as shown in Eq.~\ref{eq:1}. %These variables form the basis of the fit used in the $B^{+}\!\to K^{+}\nu\bar{\nu}$ analysis. 
The expected event yield in a given point $x$ is obtained from the selection efficiency and the model prediction. For a new model, we reweight the SM events in each $q^2$ bin by the ratio of the new-model prediction to the SM one, $\frac{\sigma_1}{\sigma_0}$, which provides the predicted distribution for that model. 

\begin{equation}\label{eq:1}
n_{1}(x)=\sum_{q^{2}\rm{bins}}
n_{0,q^{2}}(x)\left[\frac{\sigma_{1}(q^{2})}{\sigma_{0}(q^{2})}\right],
\qquad x\equiv \left(q^{2}_{\mathrm{rec}},\, \eta_{\mathrm{BDT_2}}\right).
\end{equation}

\begin{figure}[htbp]
    \centering 
    \includegraphics[width=0.5\linewidth]{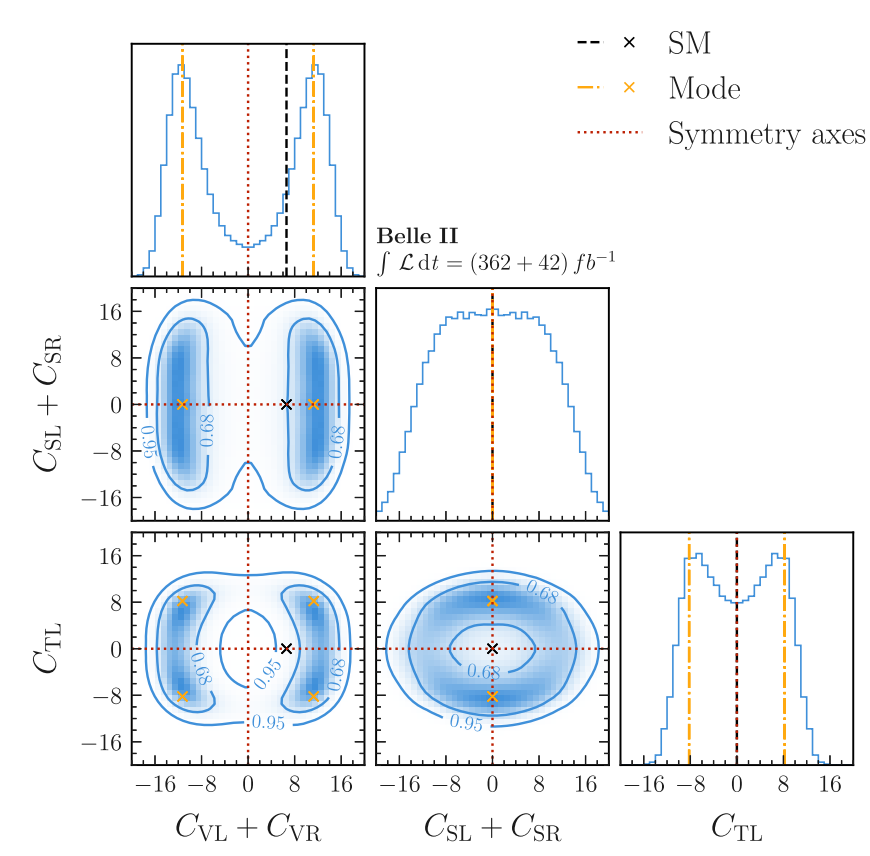}
    \caption{The marginalized posterior for the Wilson coefficients. Dashed black lines indicate the SM predictions, and the yellow lines correspond to the posterior modes. }
    \label{fig:knn3}
\end{figure}

%The Weak Effective Theory (WET) provides a unified framework to investigate possible NP effects by describing the SM and NP contributions in terms of effective operators. 
The Weak Effective Theory (WET) provides a unified framework for studying possible NP effects, as it describes contributions from both the SM and NP through effective operators. In this formalism, vector, scalar, and tensor Wilson coefficients ($C$) contribute to the differential branching fraction of the $B^{+}\!\to K^{+}\nu\bar{\nu}$ decay as predicted by the WET is given by~\cite{wet2}. 
We extract the marginal posterior distributions for the unconstrained combinations $C_{VL}+C_{VR}$, $C_{SL}+C_{SR}$, and $C_{T}$, which are treated as the parameters of interest together with the relevant hadronic form factors. The posterior maximum is located at $(11.3,\;0.0,\;8.2)$ in this parameter space. %The corresponding 95\% credible intervals are $[1.9,\,16.2]$, $[0.0,\,15.4]$, and $[0.0,\,11.2]$, respectively. The diagonal panels of the figure display the one-dimensional posterior densities, while the off-diagonal panels show two-dimensional sample densities (Fig.~\ref{fig:knn3}).
As shown in Fig.~\ref{fig:knn3}, our work obtains that an enhanced vector contribution, together with a non-zero tensor one, provide a better description of data. %As shown in Fig.~\ref{fig:knn4}~(b), the fit performed with WET-weighted events shows an improved agreement with the data compared to the SM-only unconstrained fit. Finally, we perform a fit to the $q^{2}$ spectrum using the WET-weighted events and obtain a significance of 3.3$\sigma$ for the observation of the $B^+\to K^+\nu\bar{\nu}$ signal.

\section{First $B \to X_s \nu\bar\nu$ search at Belle~II using hadronic $B$-tagging at Belle II}
\label{SNN}

The Belle II experiment search for the inclusive FCNC decay $B \to X_s \nu\bar{\nu}$ targets a precisely predicted SM branching fraction~\cite{svv_sm} of $[(3.35-3.62) \pm 0.11] \times10^{-5}$, providing a complementary probe to $B\to K\nu\bar{\nu}$. The hadronic $B$-tagging is used for tag side reconstruction, and  $X_s$ in the signal side is reconstructed in a broad set of final states containing a kaon, including $K n\pi$ ($n=0,1,2,3,4$), $3K$, and $3K\pi$ modes. This is the first time to study the channel with double $\nu$ in the final states using sum of exclusive method. 

Background suppression relies on a BDT calibrated with $B \to X_s J/\psi(\to\mu^+\mu^-)$ control samples, while normalization is constrained using off-resonance data and sidebands. A two-dimensional fit in BDT output and $X_s$ mass is performed across three regions: Region~1, enriched in $K$; Region~2, enriched in $K^{*}$; and Region~3, corresponding to the non-resonant modes. No signal is observed, Table 1 summarizes the selection efficiencies ($\epsilon$) and fit results for the three mass regions, together with the corresponding upper limits (ULs). Combining the three mass regions, an inclusive upper limit of $3.2\times10^{-4}$ at 90\% CL is set to $B \to X_s \nu\bar\nu$ for the first time.

\begin{table}[h!]
\centering
    \caption{$B \to X_s \nu\bar\nu$ fit results in three mass regions.}
\begin{tabular}{c c c c c c c}
\hline\hline
$M_{X_s}\ [\mathrm{GeV}/c^2]$ & $\epsilon\ [10^{-3}]$ & $N_{\rm sig}$ & central value $\mathcal{B}[10^{-5}]$ &  UL$_{\rm obs}[10^{-5}]$ & UL$_{\rm exp}[10^{-5}]$\\
%\multicolumn{3}{c}{$\mathcal{B}\ [10^{-5}]$} \\
%\cline{4-6}
 %& & & central value & UL$_{\rm obs}$ & UL$_{\rm exp}$ \\
\hline
$[0, 0.6]$ & 2.93 & $6^{+18+19}_{-17-16}$ & $0.3 \pm 0.8^{+0.9}_{-0.7}$ & 2.2 & 2.0 \\
$[0.6, 1.0]$ & 1.32 & $36^{+27+31}_{-26-26}$ & $3.5^{+2.6+3.1}_{-2.5-2.6}$ & 9.5 & 6.6 \\
$[1.0, m_B]$ & 0.62 & $24^{+44+62}_{-43-53}$ & $5.1^{+9.2+12.9}_{-8.8-11.0}$ & 31.2 & 26.7 \\
%Full range & 0.97 & $66^{+64+95}_{-62-81}$ & $8.8^{+8.5+12.6}_{-8.2-10.8}$ & 32.2 & 24.4 \\
\hline\hline
\end{tabular}
\label{tab:efficiency_results}
\end{table}

\begin{figure}[h]
    \centering 
 \includegraphics[width=0.45\linewidth]{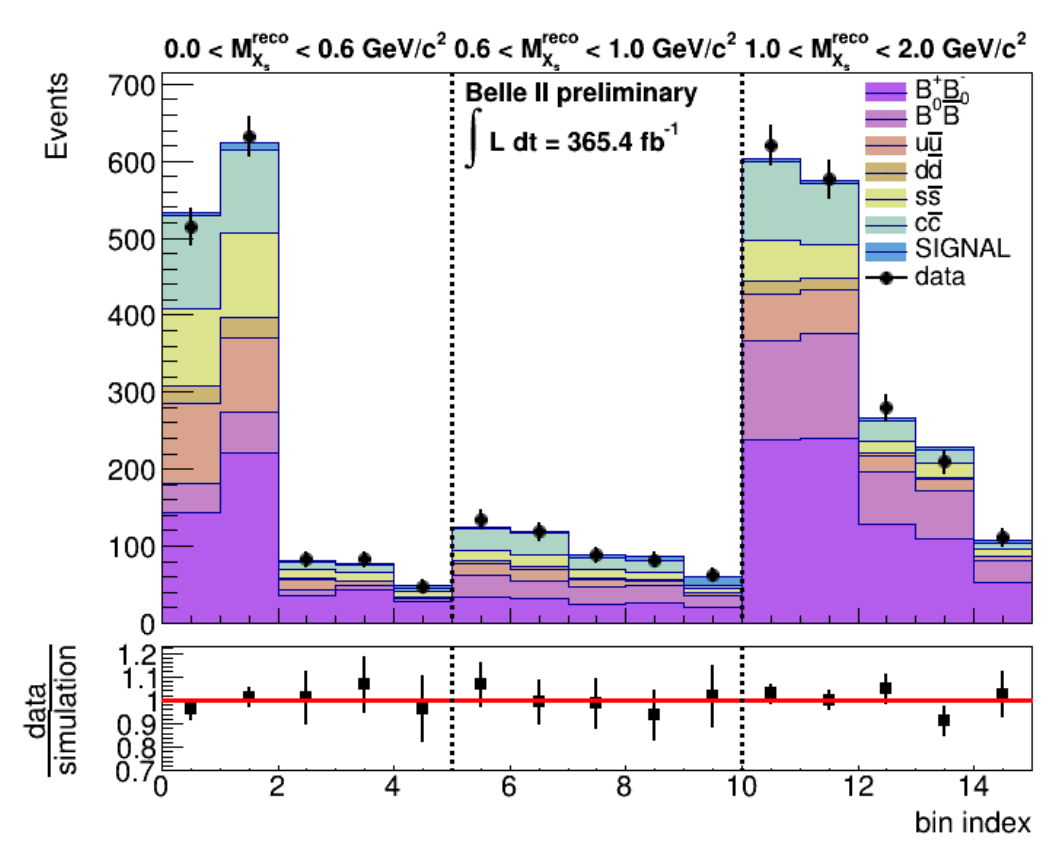}
    \caption{The bin index distribution after the fit, for data and histogram templates.}
    \label{fig:snn}
\end{figure}

\section{Conclusion}

The Belle II detector, with its large solid-angle coverage and well-constrained initial-state kinematics, provides an excellent environment for studying $B$ decays with missing energy using various tagging techniques. Current results demonstrate that Belle II can already achieve competitive—and in some cases world-leading—precision with relatively modest data samples.

We present a model-agnostic reinterpretation of the $B^{+}\to K^{+}\nu\bar{\nu}$ result~\cite{knn} within the WET framework, providing the first credible intervals for $b\to s$ Wilson coefficients from Belle~II data. By releasing the corresponding likelihood~\cite{hepdata}, this work enables rigorous tests of new-physics scenarios and establishes a framework for future Belle~II reinterpretations.

In addition, using the same dataset and hadronic $B$-tagging, we study the FCNC processes $B \to X_s \nu\bar{\nu}$ and $B^0 \to K^{*0}\tau^+\tau^-$. We set 90\% CL upper limits of $3.2 \times 10^{-4}$ and $1.8 \times 10^{-3}$, respectively, with the former representing the most stringent constraint to date.

Furthermore, by combining Belle and Belle II datasets, we perform the first search for $B^0 \to K_S^{0}\tau^+\tau^-$ and a search for $B^+\to K^+ \tau^+\tau^-$ using hadronic $B$-tagging. We obtain the most stringent upper limits of $8.4 \times 10^{-4}$ and $5.6 \times 10^{-4}$ at 90\% CL, respectively.

\section*{References}
%\bibliography{moriond}

\end{document}